\documentclass[conference]{IEEEtran}
\usepackage{flushend}

\usepackage{ifpdf}

\usepackage{cite}

\ifCLASSINFOpdf
  \usepackage[pdftex]{graphicx}
\else
  \usepackage[dvips]{graphicx}
  \DeclareGraphicsExtensions{.eps}
\fi
\begin{document}
%
\title{Experimental Study of a Triode Reflex Geometry Vircator }

\author{\IEEEauthorblockN{Vladimir Baryshevsky, Alexandra Gurinovich, Evgeny Gurnevich and Pavel Molchanov}
\IEEEauthorblockA{Research Institute for Nuclear Problems, Belarusian State University, Minsk, Belarus\\
\& Electrophysical Laboratory,
Minsk, Belarus\\
Email: gur@inp.bsu.by}}


%


\maketitle

\begin{abstract}

Triode reflex geometry vircator operating within 3.0 - 4.2~GHz
range with efficiency up to 6$\%$ is developed and experimentally
investigated.
Shiftable reflectors are shown to enable frequency tuning and
output power control.
Radiation frequency and power are analyzed for different
cathode-anode gap values and varied reflector positions.

\end{abstract}


%
\IEEEpeerreviewmaketitle

\section{Introduction}
Microwave devices with a virtual cathode in a triode reflex
geometry attract many researches by the ability to operate without
a guiding magnetic field, enhanced tunability of the operation
frequency and high output power \cite{8,9,10,11,22,23,24}.
The reflex triode vircator modified with the inclusion of
reflecting strips \cite{9} provided microwave peak power output as
high as 330 MW at 11$\%$ efficiency.

In the present paper the triode reflex geometry vircator with two
shiftable reflectors (disc-shaped and reflecting stripes) is
experimentally studied. Radiation frequency and power are analyzed
for different cathode-anode gap values and varied reflector
positions.


\section{System description}

\subsection{Pulsed power supply}

The developed triode reflex geometry vircator is driven by a
pulsed power supply, similar to that in \cite{19}, using a
30~kJ/100~kV capacitor bank and an exploding wire array (EWA)
capable of generating a 600~kV voltage pulse.

A high-voltage pulse of positive polarity is applied to the
centered in the vacuum chamber anode of 200~mm diameter.
The EWA consists of parallel connected oxygen-free
high-conductivity (OFHC) 99.99~$\%$ purity copper wires 100~$\mu$m
in diameter. The length and the number of wires can vary to match
the EWA and vircator impedances. The EWA case is designed to be
filled with gas (nitrogen or nitrogen and $SF_6$ mixture) at
pressures up to 0.6~MPa, but the array can also be
fired in air.
A pressurized $SF_6$ spark gap sharpens the high-voltage output,
so that the diode voltage pulse approaching 460~kV with a rise
time well below 100~ns is generated: the gas pressure and the gap
between the electrodes can be varied.

\begin{figure}
  \begin{center}
  \includegraphics[width=7.0 cm]{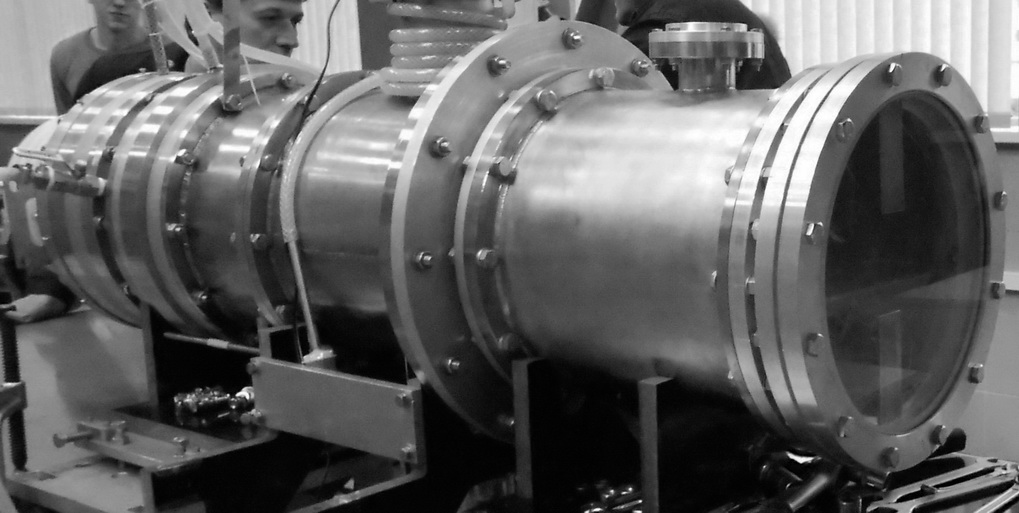}
  \end{center}
 \caption{System photo}\label{fig:1}
\end{figure}

\begin{figure}
\centerline{\includegraphics[width=8 cm]{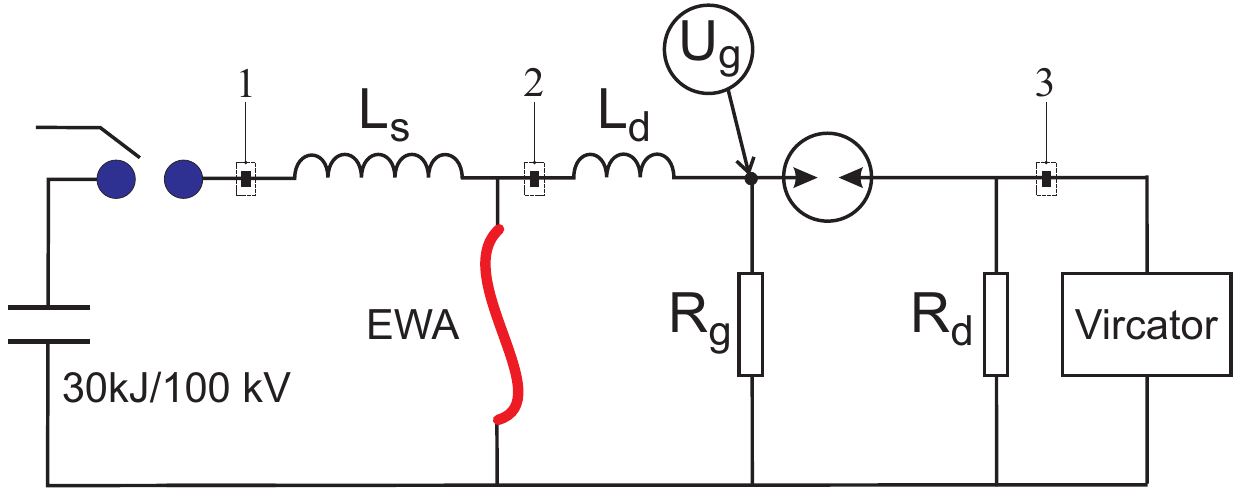}}
 \caption{Equivalent electrical scheme of the system: marks 1, 2 and 3 indicate location of three Rogowski coils.}\label{fig:2}
\end{figure}

The equivalent electrical scheme of the system (see
Fig.~\ref{fig:1}) is shown in Fig.~\ref{fig:2}.

\subsection{Triode reflex geometry vircator}

The vacuum chamber of 300~mm diameter and 600~mm length encloses
the triode reflex geometry vircator operating in the frequency
range from 3.0 to 4.2~GHz (see Fig.~\ref{fig:3} -
Fig.~\ref{fig:5}).

{\begin{figure}[h]
   \centerline{\includegraphics[width=8.0 cm]{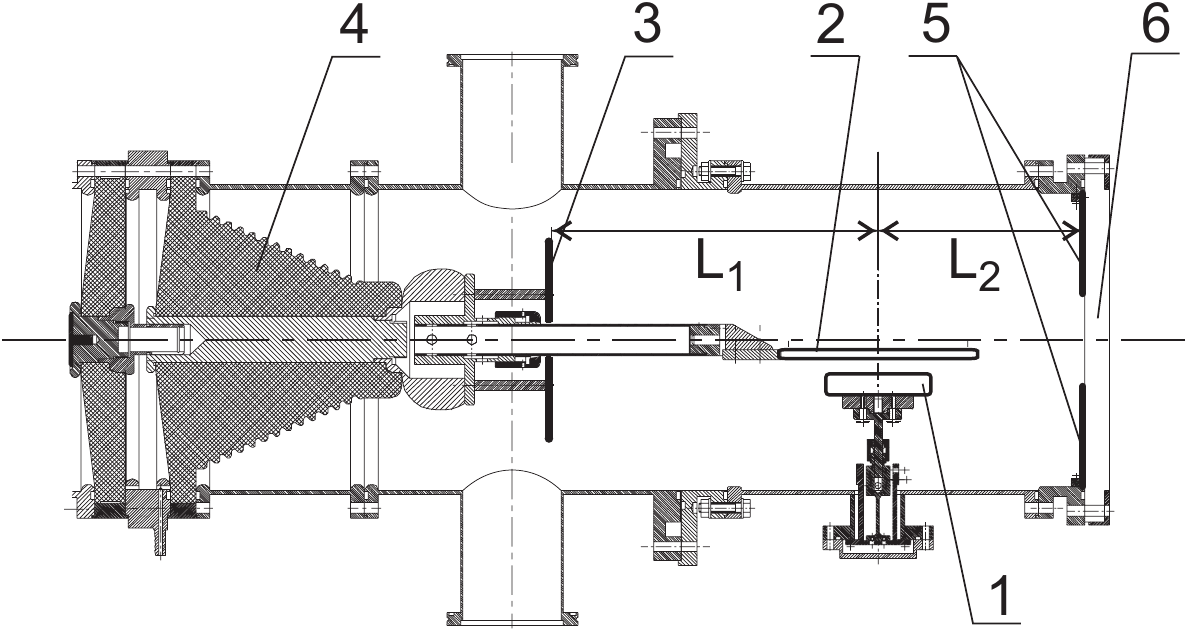}}
  \caption{Resonator geometry: 1 -- the explosive-emission cathode, 2 -- anode mesh and anode holder centered in the vacuum chamber, 3 -- disk-shaped reflector, 4 -- high-voltage vacuum feed-through,
  5 -- output shiftable reflector with rectangular brass stripes, 6 -- output window}\label{fig:3}
\end{figure}
Shown in Fig.~\ref{fig:3},  $L_1$ and $L_2$ are the variable
distances from the cathode axis to the disk-shaped and output
shiftable reflectors, respectively.

\begin{figure}[ht]
\begin{center}
\includegraphics[height=4 cm]{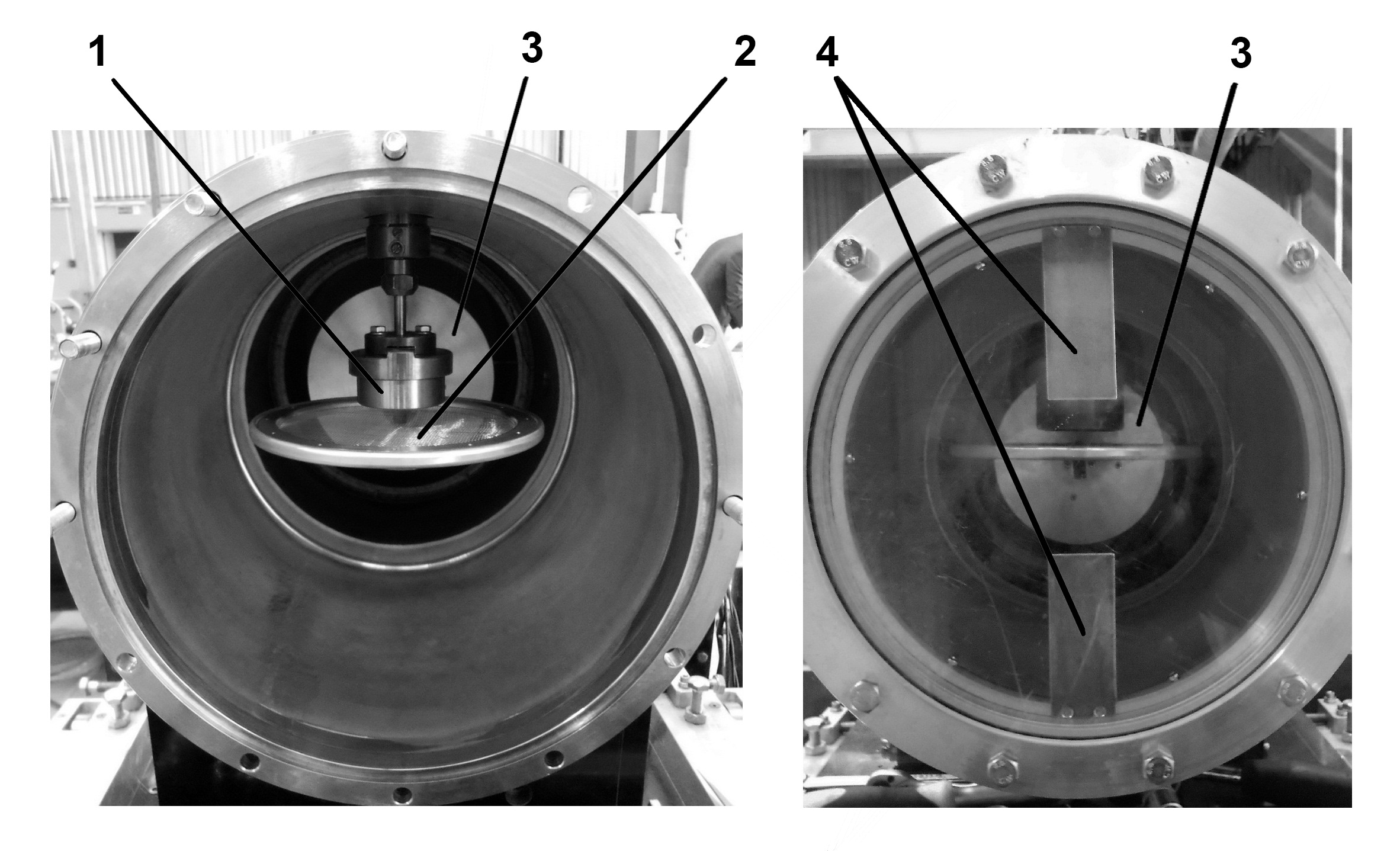}

\end{center}
\caption{Resonator photos: 1 -- the explosive-emission cathode,
  2 -- anode,
  3 -- disk-shaped reflector,
  4 -- output shiftable reflector with rectangular brass stripes} \label{fig:4}
\end{figure}

Stainless steel woven anode mesh with 77~$\%$ geometric
transparency is used in the experiments; the diameter of the mesh
thread is 224~$\mu$m.
Solid type cathode of 60~mm diameter with surface hatching is made
of dense fine-grained graphite MPG-8 (produced by NIIGraphite,
Moscow, Russia). The output reflector consists of two rectangular
brass stripes 40~mm in width and 100~mm in length, housed at a
variable distance from the output window, normally to the anode
plane position.

The value of the cathode-anode gap can be fixed with
0.1~mm accuracy. Coplanarity of cathode and anode surfaces is controlled.
The cathode-anode gap value and reflector positions (defined by
$L_1$ and $L_2$) can be tuned to provide stable single frequency
generation and the highest output power.

\begin{figure}[ht]
\begin{center}
\includegraphics[width=3.1 cm]{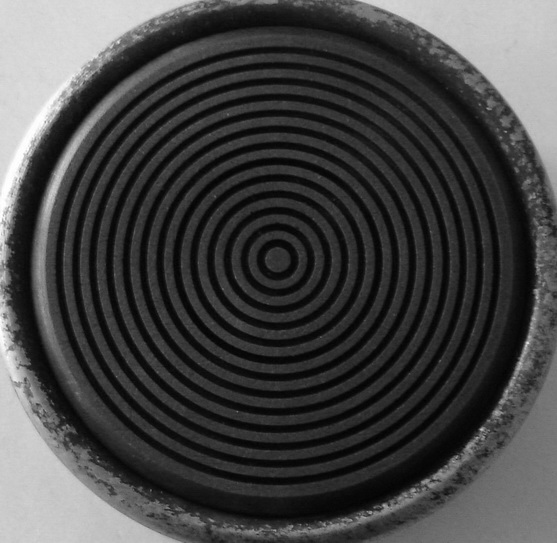}
\includegraphics[width=5.3 cm]{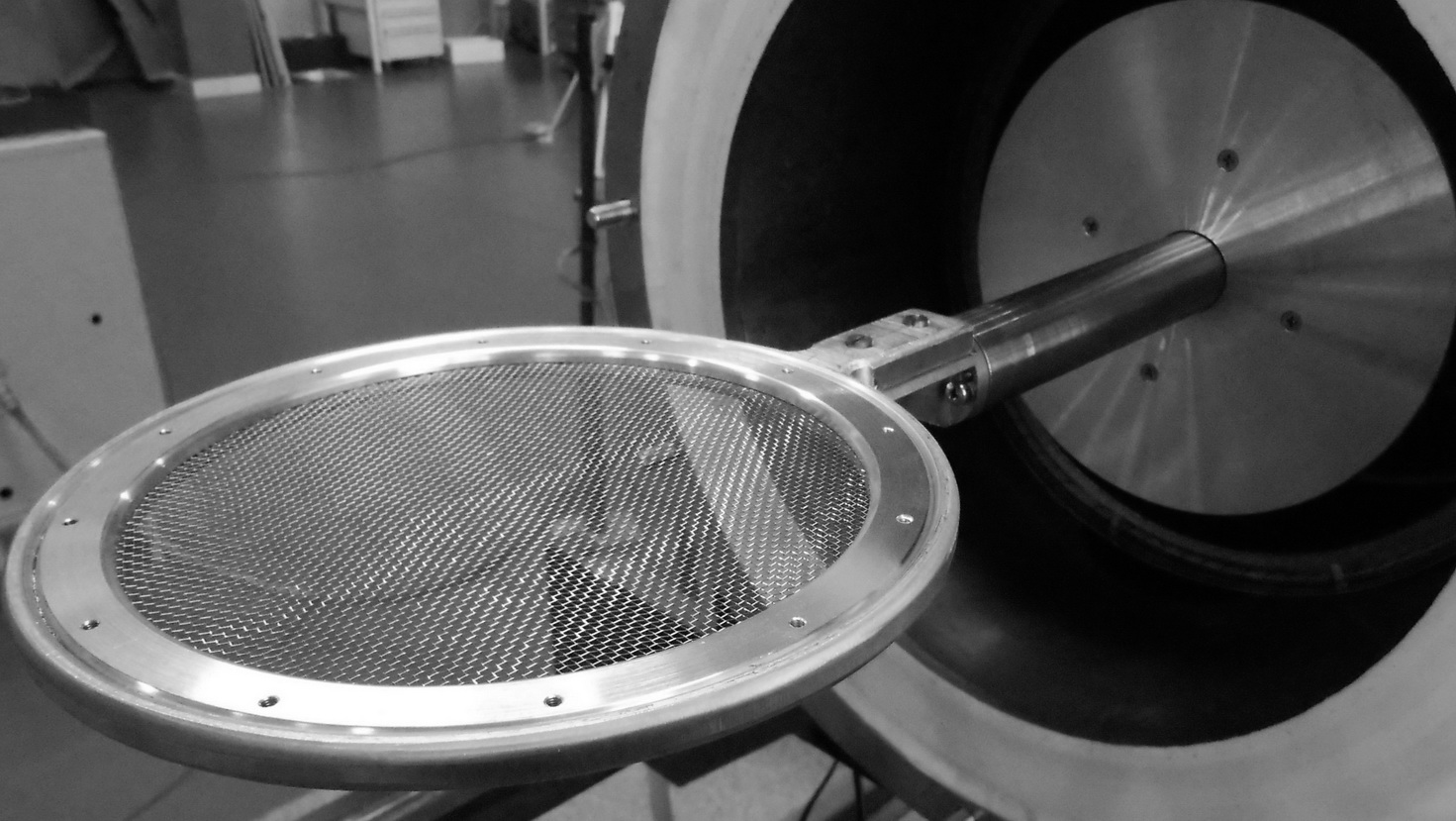}
\end{center}
\caption{Left: cathode. Right: anode mesh, its holder and
disk-shaped reflector.} \label{fig:5}
\end{figure}

\subsection{Diagnostic equipment}

Three Rogowski coils enable detecting the current derivatives in
the system at points 1, 2, and 3 in Fig.~\ref{fig:2}
Measured inductances and resistances in the circuit allow
evaluating the EWA voltage and the voltage applied to the cathode.
The microwave emission is detected at a distance of 11.5~m from
the output window in the main lobe direction by two receiving
antennas Geozondas 1-4.5GHz and Tektronix oscilloscopes TDS7704
and TDS7354.
All the measuring channels are synchronized.

\section{Experimental results}

The experimental investigation of a triode reflex geometry
vircator was intended to ascertain the system parameters providing
stable, single frequency, and high-power microwave generation.
System operation was analyzed for different $L_1$ and $L_2$ values
and cathode-anode gaps.
Experimental results obtained for 3 cases are presented:

1. $L_1=290$~mm and $L_2=188$~mm with cathode anode gap values
from 16 to 20~mm;

2. $L_1=290$~mm and $L_2=164$~mm with cathode anode gap values
from 16 to 18~mm;

3. $L_1=290$~mm with removed reflecting stripes at 16~mm
cathode-anode gap 16~mm.

The typical voltage and current signals obtained for EWA
containing 21 wires of 750~mm length are presented in
Fig.~\ref{fig:6}. Two peaks emerged on the voltage curve: the
first one (left) corresponded to spark gap closing and the second
(right) marked the maximum diode voltage. The maximum diode
voltage was as high as 430~kV and the amplitude of electron beam
current was about 17~kA.

\begin{figure}[th]
\centering
\includegraphics[width=8 cm]{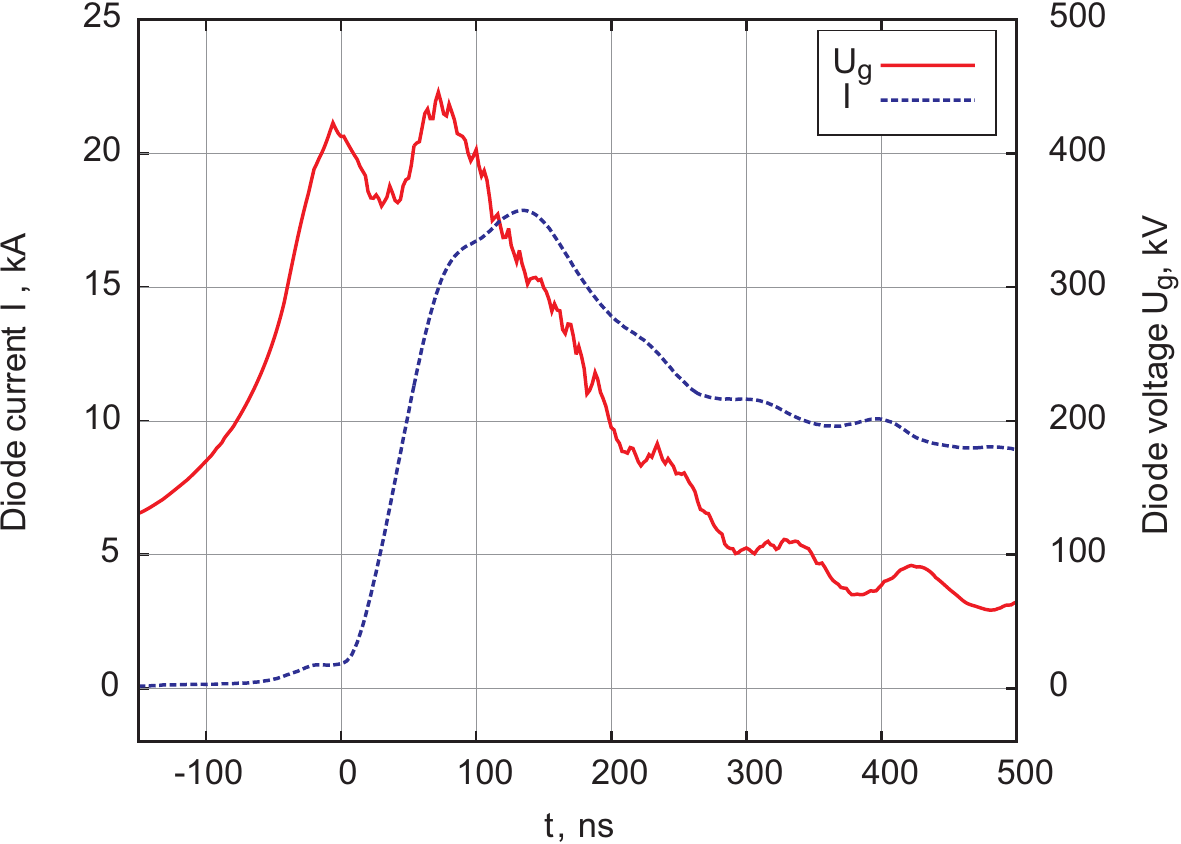}
\caption{Evaluated voltage $U_g$ and diode current $I$ for
vircator with reflector positions $L_1$=290~mm and $L_2$=188~mm
and cathode-anode gap 17~mm} \label{fig:6}
\end{figure}

Knowing the gain and directivity of the system output and the
parameters of the receiving antenna, we can convert the electric
field strength measured at a certain distance from the output
window to the radiated power.

The maximum radiated power for reflector positions $L_1=290$~mm
and $L2=188$~mm was observed at cathode-anode gap 17~mm.
The detected microwave signal and its spectral content are shown
in Fig.~\ref{fig:7} - Fig.~\ref{fig:8}: frequency, peak power and
maximal electric field strength at 11.5~m distance from the output
window measured 3.4~GHz , 400~MW and 55~kV/m, respectively.
Radiation spectra shown in Fig.~\ref{fig:8} were obtained in
different shots at similar system parameters. Two close
frequencies with different spectral power were detected: 3.37~GHz
stronger line and much weaker 3.13~GHz one. Spectra a) and b) in
Fig.~\ref{fig:8} demonstrate maximal and minimal observed
difference in spectral power for these two lines.

\begin{figure}[th]
\centering
\includegraphics[width=7 cm]{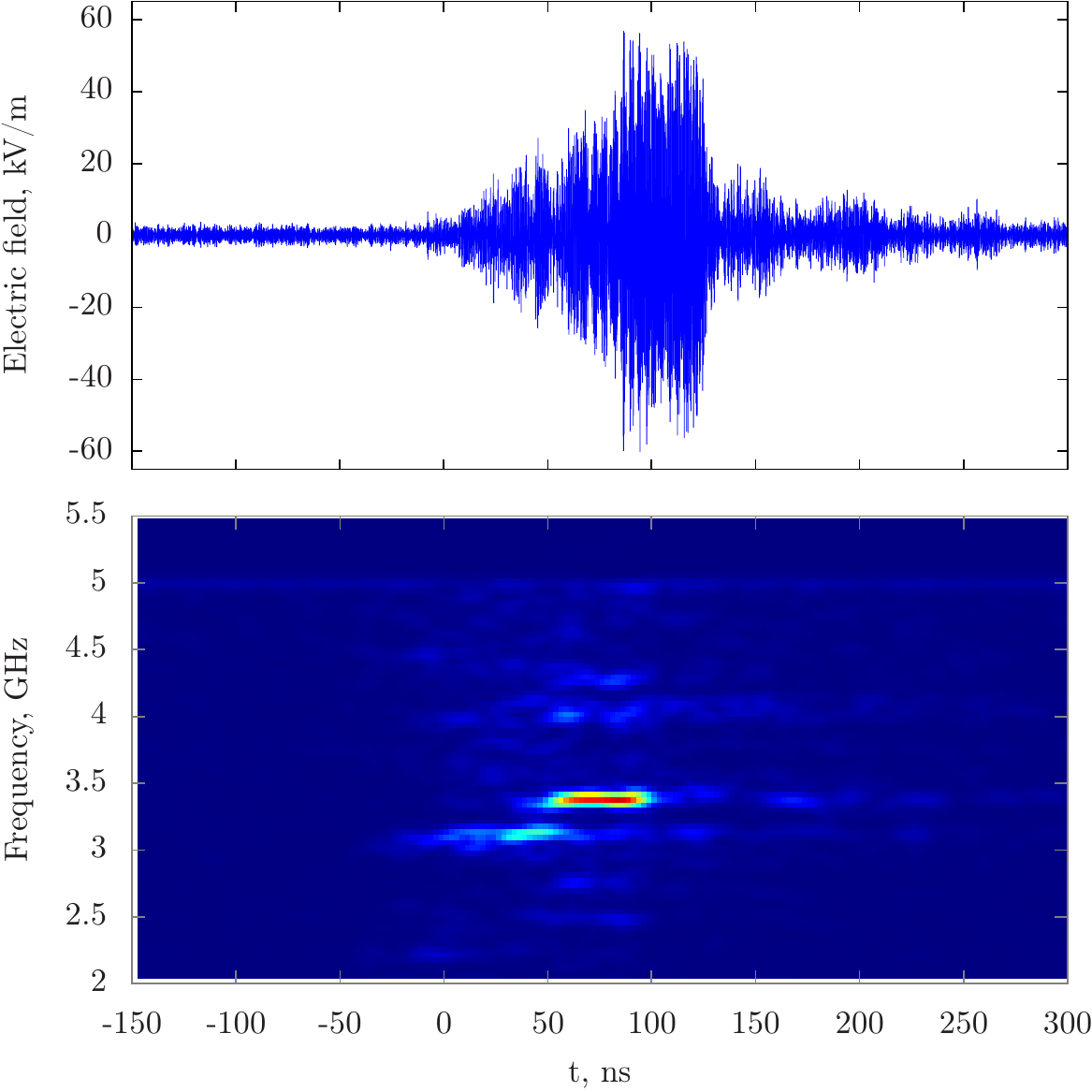}
\caption{Detected microwave signal and its spectral content
obtained for reflector positions $L_1$=290~mm and $L_2$=188~mm and
cathode-anode gap 17~mm} \label{fig:7}
\end{figure}

Variation of the cathode-anode gap value over the range 20 to
17~mm demonstrated smooth change of the radiation frequency within
the range 3.0 to 3.37~GHz (see Fig.~\ref{fig:9}).
In each diagram zone, which corresponds to the fixed cathode-anode
gap value, the sum of frequency spectrums obtained in several
experiments under similar conditions is presented.
Change of the cathode anode gap value from 17 to 16~mm at fixed
position of the output reflector lead to the hop of the radiation
frequency from 3.37 to 4.16~GHz, which is also shown in
Fig.~\ref{fig:9}.

\begin{figure}[th]
\centering
\includegraphics[height=3.2 cm]{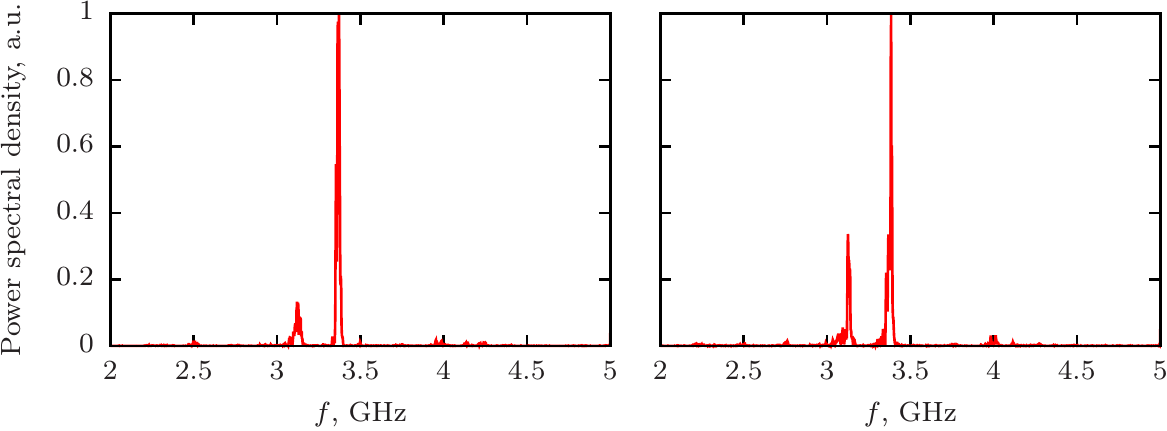}
\caption{Radiation spectra for reflector positions $L_1$=290~mm
and $L_2$=188~mm and cathode-anode gap 17~mm} \label{fig:8}
\end{figure}

\begin{figure}[th]
\centering
\includegraphics[width=7 cm]{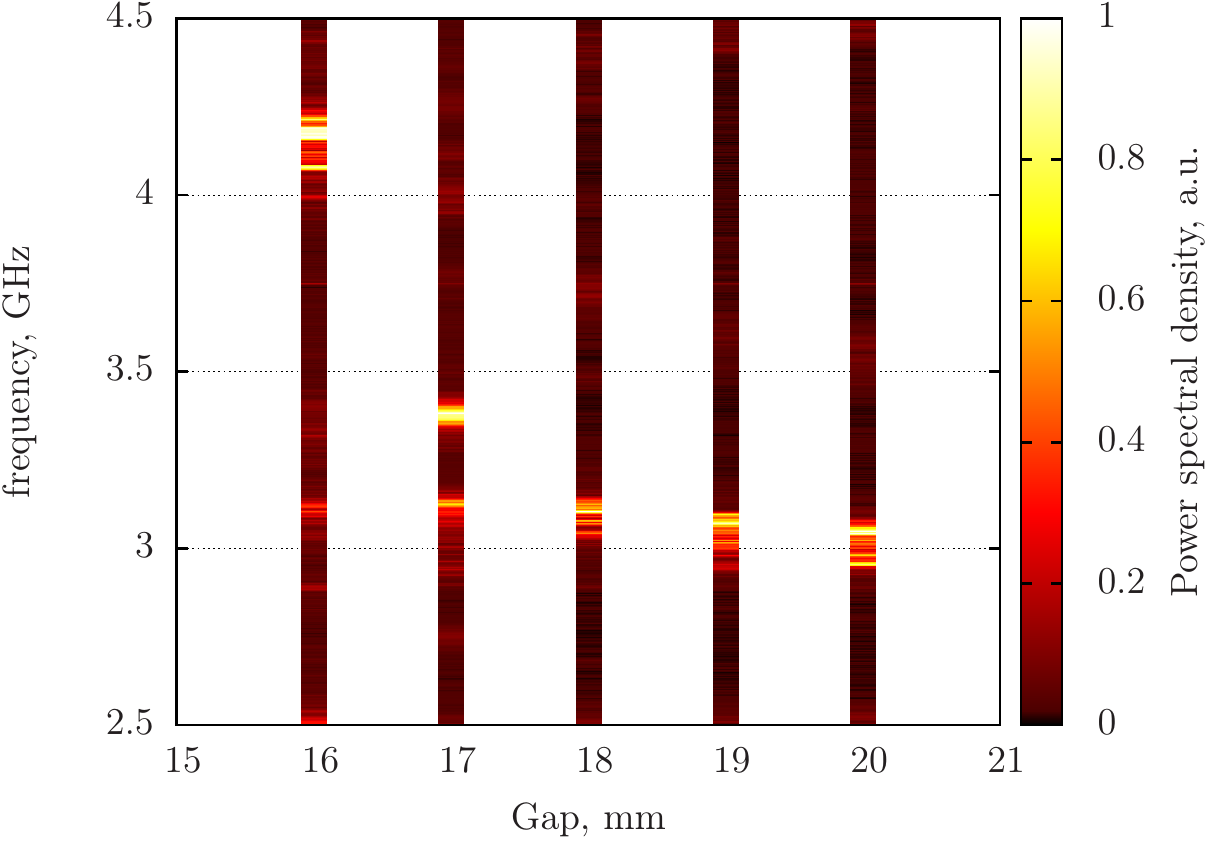}
\caption{Comparison of spectra obtained for different
cathode-anode gap values for reflector positions $L_1=290$~mm and
$L2=188$~mm.}\label{fig:9}
\end{figure}

The maximum radiated power in the triode reflex geometry vircator
with reflector positions  $L_1=290$~mm and $L_2=164$~mm was
observed at cathode-anode gap 16~mm.
About 460~MW at 4.16~GHz frequency was produced at maximum diode
voltage as high as 460~kV and amplitude of electron beam current
18~kA (see Fig.~\ref{fig:10}).  Detected microwave signal with
electric field strength amplitude $ \sim 70$~kV/m and its spectral
content are shown in Fig.~\ref{fig:11} - Fig.~\ref{fig:12}.
Cathode-anode gap change to 17 and 18~mm resulted in spectrum
broadening and shift of its central frequency. Detected microwave
signal and its spectral content obtained in experiments with
reflector positions $L_1$=290~mm and $L_2$=164~mm at cathode-anode
gap 18~mm are shown in Fig.~\ref{fig:14}.

\begin{figure}[h]
\centering
\includegraphics[width=7 cm]{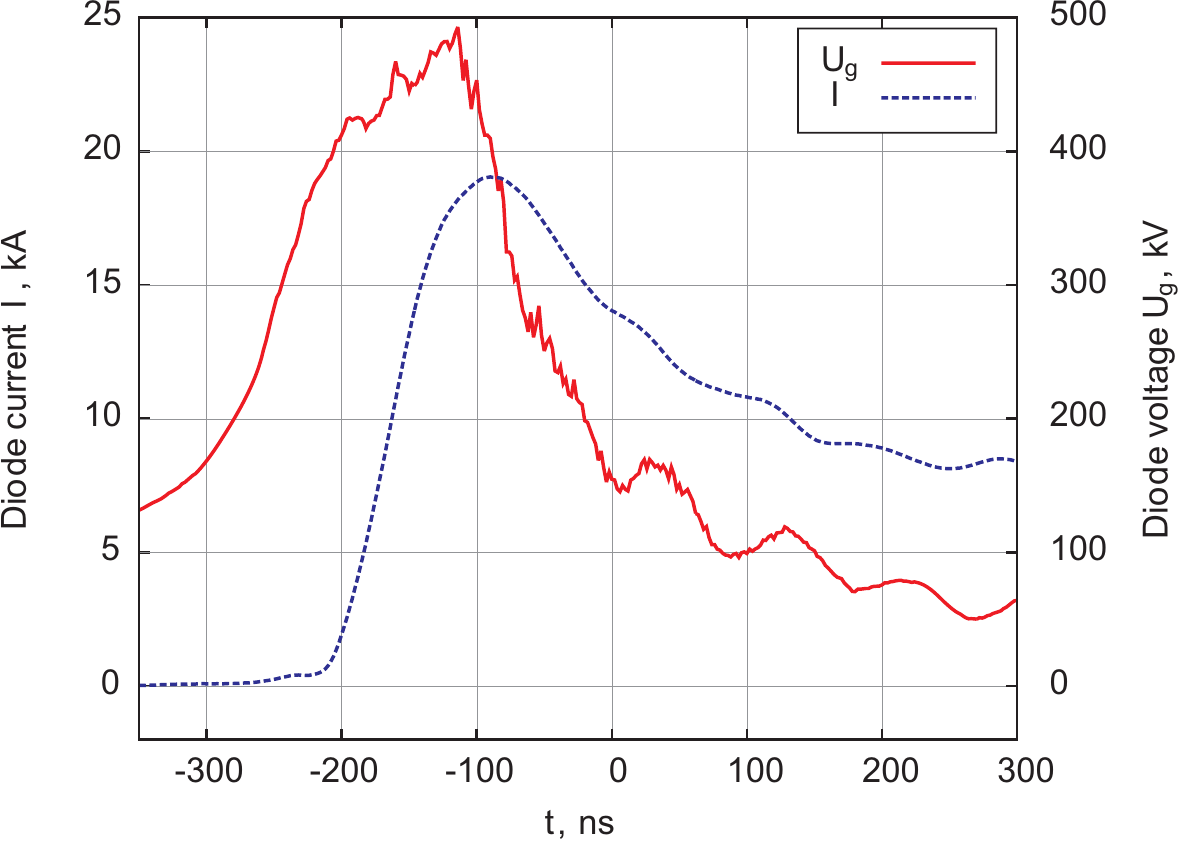}
\caption{Evaluated voltage $U_g$ and diode current $I$ for
vircator with reflector positions $L_1$=290~mm and $L_2$=164~mm at
cathode-anode gap 16~mm} \label{fig:10}
\end{figure}

\begin{figure}[h]
\centering
\includegraphics[width=8 cm]{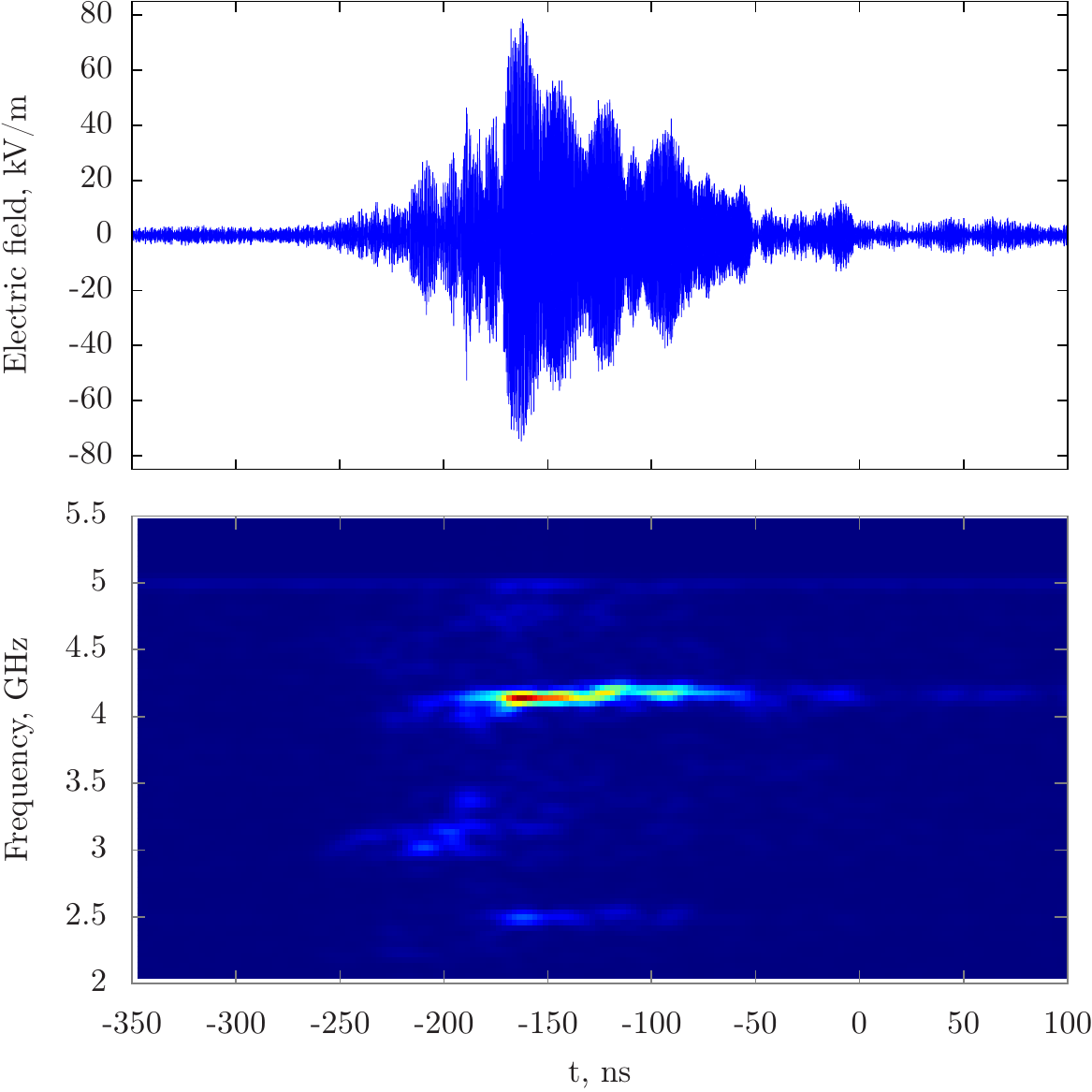}
\caption{Detected microwave signal with the highest amplitude and
its spectrogram for vircator with reflector positions $L_1$=290~mm
and $L_2$=164~mm at cathode-anode gap 16~mm} \label{fig:11}
\end{figure}

\begin{figure}[h]
\centering
\includegraphics[width=8 cm]{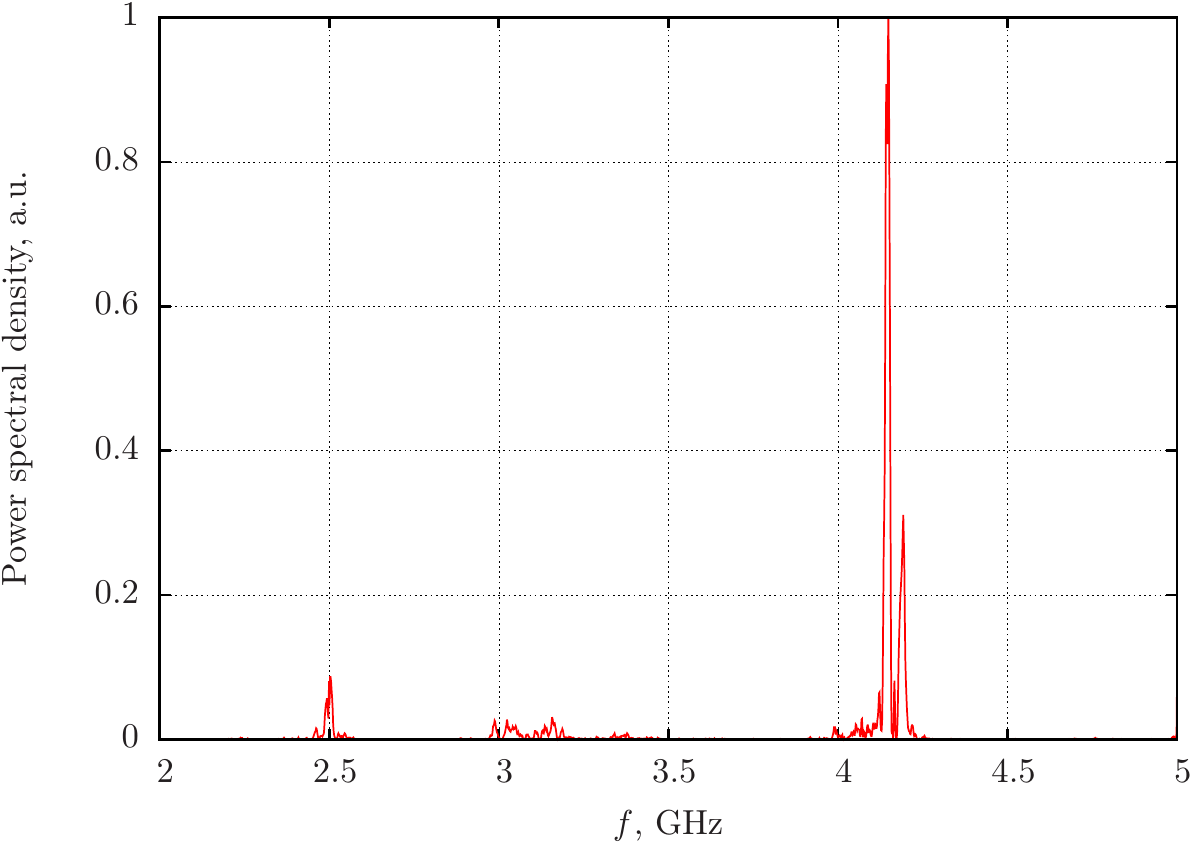}
\caption{Radiation spectrum for vircator with reflector positions
$L_1$=290~mm and $L_2$=164~mm at cathode-anode gap 16~mm}
\label{fig:12}
\end{figure}

\begin{figure}[h]
\centering
\includegraphics[width=8 cm]{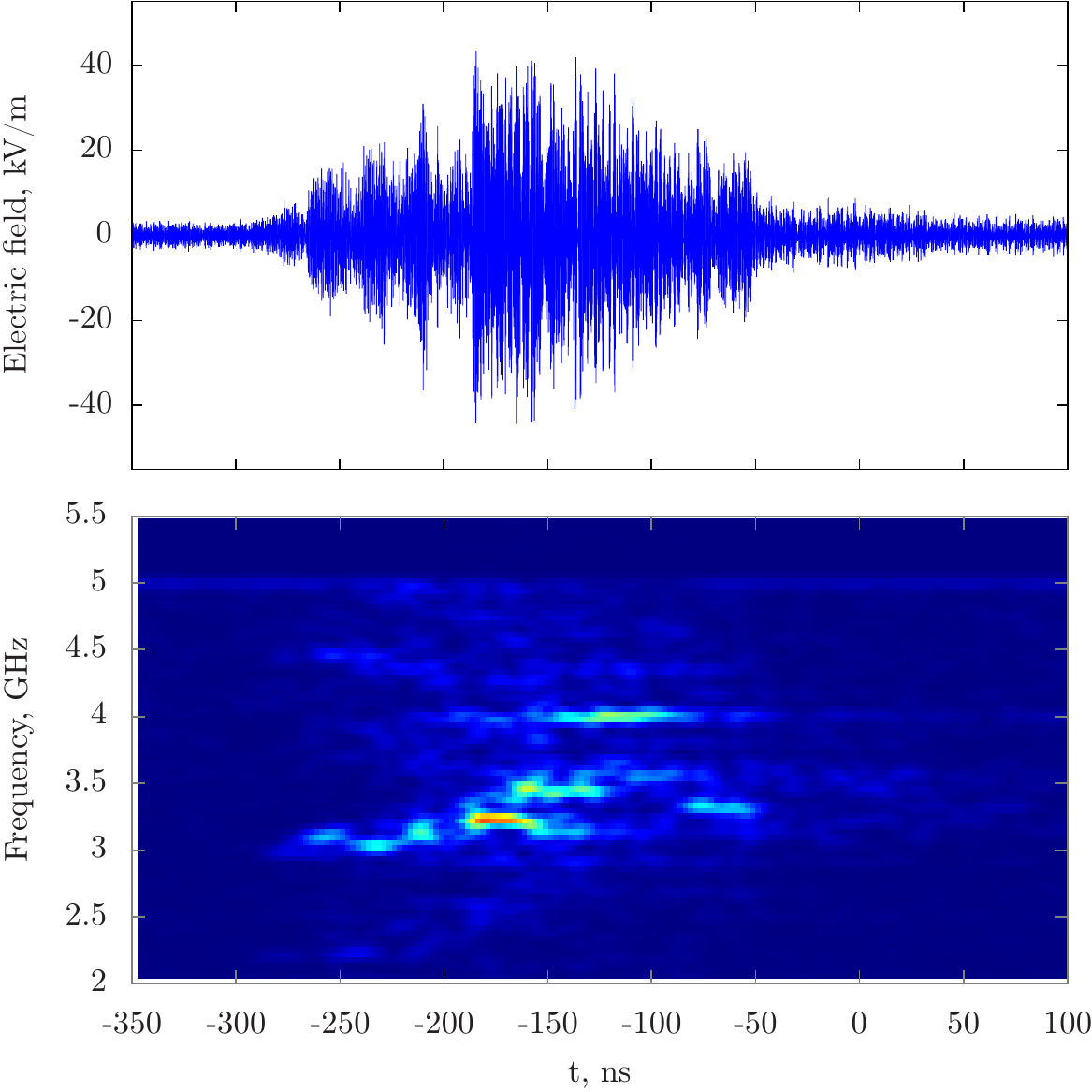}
\caption{Detected microwave signal and its spectral content
obtained in experiments with reflector positions $L_1$=290~mm and
$L_2$=164~mm at cathode-anode gap 18 mm} \label{fig:14}
\end{figure}

Triode reflex geometry vircator was also studied without shiftable
output reflector. Reflecting stripes of the output reflector were
dismounted, while the disk-shaped reflector was fixed at the
position with $L_1$=290 mm.
The  complicated multifrequency  generation over the range 3.0 to
4.2~GHz obtained in the experiment is shown in Fig.~\ref{fig:15}.
The electric field strength value measured at a 11.5 m distance
from the output window in the main lobe direction in the series of
experiments did not exceed 35 kV/m  (see Fig.~\ref{fig:15}).

\begin{figure}[th]
\centering
\includegraphics[width=8 cm]{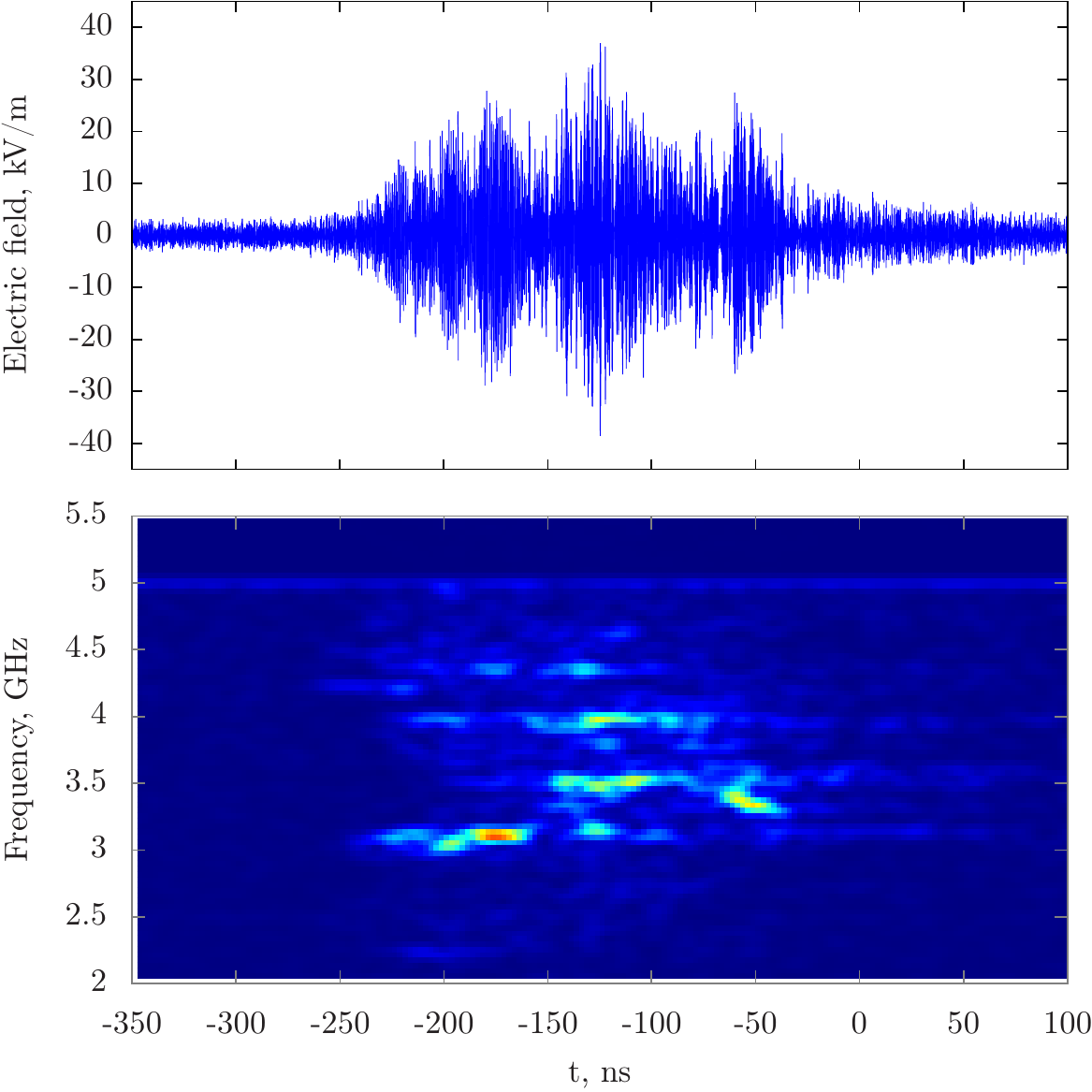}
\caption{Detected microwave signal and its spectral content
obtained in experiments without output reflector and cathode-anode
gap value of 16 mm} \label{fig:15}
\end{figure}

\section{Conclusion}
Triode reflex geometry vircator operating within 3.0 - 3.37~GHz
and 4.0 - 4.2~GHz ranges with efficiency up to 6$\%$ was developed
and experimentally investigated.
Shiftable reflectors were shown to enable frequency tuning and
output power control.
Radiation frequency and power were analyzed for different
cathode-anode gap values and varied reflector positions. The
highest radiation power of 400 and 460~MW was obtained in the
experiments with radiation frequency 3.37 and 4.16~GHz,
respectively.

Significant influence of the output reflector on both the
radiation efficiency and frequency is shown. Application of the
output reflector enabled to achieve 1.6 times higher electric
field strength value in the main lobe direction as compared to the
case when reflector was absent.
The influence of the cathode-anode gap value is also considered.


\section*{Acknowledgment}

The authors would like to thank Victor Evdokimov for the
sophisticated system design and Nikolai Belous for permanent
involvement in the experimental activities.




\begin{thebibliography}{1}

\bibitem{8} A.N.~Didenko, \emph{Generation of high power RF pulses in magnetron and reflex triode systems},
High Power Electron and Ion Beam Research and Technology, Proc.
III Intern. Conf., Novosibirsk, 1979, V. 2, pp. 683-691.

\bibitem{9} J.J.~Mankowski, Xupeng Chen, J.C.~Dickens, Magne Kristiansen,
\emph{Experimental Optimization of a Reflex Triode Virtual Cathode
Oscillator}, High-Power Particle Beams, BEAMS 2004, pp. 426 - 429.


\bibitem{10} W.~Jiang, N.~Shimada, S.D.~Prasad, K.~Yatsui, \emph{Experimental
and Simulation Studies of New Configuration of Virtual Cathode
Oscillator},IEEE Trans. Plasma Sci., vol. 32, No. 1, 2004, pp.
54-59.

\bibitem{11} L.~Liu, L.M.~Li, X.P.~Zhang, J.C.~Wen, H.~Wan, Y.Z.~Zhang,
\emph{Efficiency enhancement of reflex triode virtual cathode
oscillator using the carbon fiber cathode}, IEEE Trans. Plasma
Sci., vol. 35, no. 2, pp. 361-368, 2007.




\bibitem{22}
V.P.~Grigoryiev, A.G.~Zherlitsyn, T.V.~Koval, G.V.~Melnikov,
P.Ya.~Isakov, \emph{Mode Structure Research of a Field in the
Triode with the Virtual Cathode with an Active Feedback},Proc.
14th Symposium on High Current Electronics,Tomsk, Russia,Izvestiya
Vuzov, Physics, 2006, V.40, no.11, pp. 372-375.

\bibitem{23}
A.G.~Zherlitsyn, G.V.~Mel'nikov,P.Ya.~Isakov, \emph{Effect of
Feedback on the Microwave Radiation in a Triode with a Virtual
Cathode},Journal of Communications Technology and Electronics,
2007, Vol. 52, No. 7, pp. 798-802.


\bibitem{24}
Y.~Chen, J.~Mankowski, J.~Walter, M.~Kristiansen, \emph{Cathode
and Anode Optimization in a Virtual Cathode Oscillator},IEEE
Transactions on Dielectrics and Electrical Insulation, Vol.14, Is.
4,  pp. 1037 - 1044, 2007.

\bibitem{19}
V.~Baryshevsky, A.~Gurinovich, E.~Gurnevich, P.~Molchanov.
\emph{Experimental Study of an Axial Vircator with Resonant
Cavity}, IEEE Transactions on Plasma Science, v. 43, No. 10, pp.
3507-3511, 2015.

\end{thebibliography}
%

\end{document}